\documentclass[%
 reprint,
superscriptaddress,
nofootinbib,
 amsmath,amssymb,
 aps,
]{revtex4-1}

\usepackage{graphicx}
\usepackage{dcolumn}
\usepackage{bm}
\usepackage{footnote}
\usepackage{ wasysym }

\usepackage[usenames,dvipsnames]{color}
\usepackage[colorlinks=true, linkcolor=BrickRed, citecolor=Blue, urlcolor=Blue, filecolor=Blue, draft]{hyperref}

\begin{document}

\preprint{}

\title{Understanding the $\gamma$-ray emission from the globular cluster 47 Tuc: evidence for dark matter?}

\author{Anthony M. Brown}\email{anthony.brown@durham.ac.uk}
\affiliation{Centre for Advanced Instrumentation, Department of Physics, University of Durham, South Road, Durham, DH1 3LE, UK}

\author{Thomas Lacroix}
\affiliation{Laboratoire Univers \& Particules de Montpellier (LUPM), Universit\'{e} de Montpellier, CNRS, Universit\'{e} de Montpellier, Montpellier, France}

\author{Sheridan Lloyd}
\affiliation{Centre for Advanced Instrumentation, Department of Physics, University of Durham, South Road, Durham, DH1 3LE, UK}

\author{C\'eline B\oe hm}
\affiliation{School of Physics, University of Sydney, Camperdown, NSW 2006, Australia}
\affiliation{Institute for Particle Physics Phenomenology, Durham University, South Road, Durham, DH1 3LE, United Kingdom}
\affiliation{LAPTH, U. de Savoie, CNRS,  BP 110, 74941 Annecy-Le-Vieux, France}
\affiliation{Perimeter Institute, 31 Caroline St N., Waterloo Ontario, Canada N2L 2Y5}

\author{Paula Chadwick}
\affiliation{Centre for Advanced Instrumentation, Department of Physics, University of Durham, South Road, Durham, DH1 3LE, UK}


\date{\today}

\begin{abstract}
47 Tuc was the first globular cluster observed to be $\gamma$-ray bright, with the $\gamma$-rays being attributed to a population of unresolved millisecond pulsars (MSPs). Recent kinematic data, combined with detailed simulations, appears to be consistent with the presence of an intermediate mass black hole (IMBH) at the centre of 47 Tuc. Building upon this, we analyse 9 years of \textit{Fermi}-LAT observations to study the spectral properties of 47 Tuc with unprecedented accuracy and sensitivity. This 9-year $\gamma$-ray spectrum shows that 47 Tuc's $\gamma$-ray flux cannot be explained by MSPs alone, due to a systematic discrepancy between the predicted and observed flux. Rather, we find a significant preference (TS $=40$) for describing 47 Tuc's spectrum with a two source population model, consisting of an ensemble of MSPs and annihilating dark matter (DM) with an enhanced density around the IMBH, when compared to an MSP-only explanation. The best-fit DM mass of 34 GeV is essentially the same as the best-fit DM explanation for the Galactic centre ``excess" when assuming DM annihilation into $b\bar{b}$ quarks. Our work constitutes the first possible evidence of dark matter within a globular cluster.  
\end{abstract}

\pacs{Valid PACS appear here}
\maketitle


\section{\label{sec:Intro}Introduction}

With ages on the order of $\sim10^{10}$ years, globular clusters are believed to represent the oldest components of our Galaxy. They have been extensively studied across the electromagnetic spectrum, revealing a large number of binary systems and millisecond pulsars (MSPs). These are believed to be a by-product of the number of stars within globular clusters, upwards of $10^5$ within a 50 cubic parsec volume. This high density leads to high encounter rates between stars and a formation rate of low-mass X-ray binaries (LMXBs) that is several orders of magnitude higher than in the Milky Way itself \cite{katz,clark}. Since MSPs are believed to result from LMXBs, this large LMXB formation rate implies that globular clusters are likely to host a large population of MSPs. Individual MSPs in our local Galactic neighbourhood have been found to be $\gamma$-ray bright \cite{2pc}, so the presence of a significant MSP population within globular clusters raises the possibility of globular clusters being $\gamma$-ray bright.

Observations by the Large Area Telescope onboard the \textit{Fermi} satellite (\textit{Fermi}-LAT) early in its science mission discovered the prominent globular cluster 47 Tuc to be $\gamma$-ray bright \cite{47tucfermiscience}. Using the ephemerides of known MSPs in 47 Tuc \cite{ephems}, the \textit{Fermi}-LAT collaboration did not find any evidence of $\gamma$-ray flux pulsation and concluded that a population of unresolved MSPs was responsible for 47 Tuc's $\gamma$-ray emission. They also concluded there was no difference in `spin-down to $\gamma$-ray luminosity' conversion efficiency of 47 Tuc's MSP population compared to our local Galactic neighbourhood MSPs. 

Radio and X-ray observations of 47 Tuc have been used to place limits on the mass of a central black hole, under the assumption that it is accreting \cite{derijcke,grindlay}. Relaxing this assumption, the spatial distribution and motion of phase-resolved pulsars, combined with detailed N-body simulations, recently revealed evidence for a $\sim2300$ M$_{\astrosun}$ intermediate mass black hole (IMBH), residing within 47 Tuc \cite{imbh}. This result prompts us to consider the presence of an enhanced DM density, referred to as a spike \cite{gondolo}, around 47 Tuc's IMBH \cite{horiuchi}. This would in turn enhance a possible $\gamma$-ray signal from DM annihilation.

We note that work by Freire et al. \cite{freire}, which uses additional kinematic information in the form of the MSP jerk, finds no strong evidence for an IMBH within 47 Tuc. They assume, however, a distance of 4.69 kpc to 47 Tuc, which is among the largest published for 47 Tuc. Even at this larger distance, they still find a number of MSPs within their sample that have line-of-sight jerk values larger than those expected for the distance and gravitational potential assumed for 47 Tuc, though they discount these MSPs on the assumption that they are the result of interactions with local stars. Importantly, the authors of \cite{freire} highlight that all their results are based on an assumed gravitational potential that does not contain an IMBH. Furthermore the authors of \cite{imbh} still find evidence for an IMBH if they assume a distance to 47 Tuc of 4.5 kpc. It is worth noting that from a radio perspective, the authors of Ref.~\cite{freire} find that the MSPs within 47 Tuc exhibit very similar characteristics to the MSPs within the Galactic disc.

Motivated by the evidence of a significant dark mass within 47 Tuc, we analyze 9 years of \textit{Fermi}-LAT observations to study the spectral properties of the globular cluster 47 Tuc. We model the 9-year averaged $\gamma$-ray spectrum with both MSP and DM annihilation processes to constrain the sources of the observed $\gamma$-rays. The structure of our letter is as follows. In Sec.~\ref{Sec:Analysis} we determine 47 Tuc's $\gamma$-ray properties. In Sec.~\ref{Sec:Interpretation} we investigate two possible interpretations of 47 Tuc's spectrum: one involving a population of MSPs and the other involving both MSPs and DM annihilation products from an enhanced DM density around the IMBH. We discuss our findings in Sec.~\ref{Sec:Discussion}, and provide conclusions in Sec.~\ref{Sec:Conclusions}. 

\section{\textit{Fermi}-LAT data analysis}
\label{Sec:Analysis}

Our study considered all photon and spacecraft data taken during the first 9 years of the \textit{Fermi}-LAT science mission, from 2008 August 4 to 2017 August 4 (Mission Elapsed Time (MET) period of 239557417 [s] to 523554222 [s]). All $0.1<E_{\gamma}<100$ GeV \textsc{source (front$+$back)} events within a $15^{\circ}$ radius of interest (RoI) centred on 47 Tuc were analysed, with the size of the RoI defined by the LAT's point spread function for 0.1 GeV photons. It is worth noting that, although MSP $\gamma$-ray emission peaks at photon energies of a few GeV, we consider the entire $0.1-100$ GeV energy range to afford us a large spectral `lever-arm' for our spectral fitting. 

In accordance with \textsc{pass8} data analysis criteria, a zenith cut of $90^{\circ}$ was applied to the data to remove $\gamma$-rays originating from the Earth's atmosphere, and good time intervals were selected by applying a `\textsc{data\_qual}$>0$ \&\& \textsc{lat\_config}$==1$' filter criterion. Throughout our analysis, a \textsc{binned} likelihood analysis was employed, using the \textit{fermipy} python tool set \cite{wood}, version \textsc{v10r0p5} of the \textit{Fermi Science Tools} and the \textsc{p8r2\_source\_v6} instrument response functions. 



The model employed during our likelihood analyses consisted of discrete point-like and extended $\gamma$-ray sources, plus diffuse $\gamma$-ray emission. The diffuse $\gamma$-ray emission detected by the LAT comprises two components: the Galactic diffuse emission, which dominates along our Galactic plane, and the isotropic diffuse emission. The Galactic component of the diffuse emission was modeled with \textit{Fermi}'s gll\_iem\_v06.fit spatial map and a power-law spectral model with the normalisation free to vary. The isotropic diffuse emission was defined by \textit{Fermi's}  iso\_\textsc{P8R2}\_\textsc{SOURCE}\_\textsc{V6}.txt tabulated spectral data, with the normalisation left free to vary. The extended source included in our model was the Small Magellanic Cloud (SMC), located $2.7^{\circ}$ from 47 Tuc. We defined the SMC spectrally by a power-law, and spatially by the \textsc{smc.fits} template provided by the \textit{Fermi}-LAT collaboration. The point-like $\gamma$-ray source population within our model was initially seeded by the Third \textit{Fermi} Source Catalog (3FGL \cite{3fgl}). In particular, we took the position and spectral shapes of all 3FGL point sources within a source RoI of $25^{\circ}$ from 47 Tuc.

To confirm the accuracy of our `3FGL point $+$ extended $+$ diffuse' model description, an initial \textsc{binned} likelihood analysis was performed, starting with the \textit{fermipy} \textsc{optimise} routine. From this initial optimisation, all insignificant sources with a test statistic\footnote{The test statistic, TS, is defined as twice the difference between the log-likelihood of two different models, 2(log $\mathcal{L}_{1} -$ log $\mathcal{L}_{0})$, where $\mathcal{L}_{1}$ and $\mathcal{L}_{0}$ are defined as the maximum likelihood with and without the source in question \cite{like}. For one degree of freedom, TS$=\sigma^2$. }, TS, $<2$ or a predicted number of photons, $Npred$, $<4$ were removed from the model. Thereafter, the normalisation of all remaining point sources within $15^{\circ}$ was left free to vary. All point sources with a TS $>25$ were left to vary spectrally, as was the spectral shape of 47 Tuc. A second \textsc{binned} likelihood was then performed with the resultant model. The \textit{Fermi Science Tool} \textsc{gttsmap} was then used in conjunction with the final best-fit model from the initial two-step likelihood analysis to construct a $21^{\circ} \times 21^{\circ}$ TS map centered on 47 Tuc. This TS map was used to reveal additional point sources of $\gamma$-rays that were not accounted for in our initial model by identifying excesses with TS $>25$. For every TS $>25$ excess, a new point source was added to our model fixed to the ($\alpha_{J2000}$, $\beta_{J2000}$) of the excess location, and described by a power-law. The normalisation and spectral index of each power-law was left free to vary, and optimised individually. 


Once all sources of $\gamma$-rays within our dataset had been accounted for, a final \textsc{binned} likelihood analysis was performed to study 47 Tuc's $\gamma$-ray properties. Integrating over the 9-year data-set, 47 Tuc is found to be a bright $\gamma$-ray source with a test statistic of TS $=5719$ and energy flux of $(2.66 \pm 0.08) \times 10^{-11}$ ergs cm$^{-2}$ s$^{-1}$. Assuming isotropic emission for the observed $\gamma$-ray flux and a luminosity distance of 4.5 kpc \cite{47tucdist}, the total luminosity of 47 Tuc in the $0.1-100$ GeV energy range is $(6.45 \pm 0.19) \times 10^{34}$ ergs s$^{-1}$. Using the \textit{Fermi Science Tool} \textsc{gtfindsrc} and the final best-fit model, 47 Tuc's $\gamma$-ray emission was localised to ($\alpha_{J2000}$, $\beta_{J2000}$ $= 6.001^{\circ}$, $-72.080^{\circ}$) with a $1\sigma$ error radius of $0.008^{\circ}$. This position is consistent with 47 Tuc's 3FGL position, and lies within 47 Tuc's $0.715^{\circ}$ tidal radius. 

The data were then binned into ten logarithmically spaced bins per decade of energy, with a likelihood analysis being performed separately for each bin. For each separate likelihood fit, all sources within the model were frozen except for 47 Tuc's normalisation. For bins with a TS $<10$, a $2\sigma$ upper limit was calculated. The resulting spectrum can be seen in Figs.~1 \& 2, with all flux error bars representing a $1\sigma$ level of statistical uncertainty. The best-fit log-parabola description of 47 Tuc's spectrum has a spectral index of $\alpha = 1.63 \pm 0.04$ and a curvature of $\beta = 0.37 \pm 0.03$. While the curvature is consistent with that reported in the 3FGL, the spectral index is found to be slightly softer due to the significant emission below 0.2 GeV that the larger 9-year dataset reveals compared to 3FGL's 4-year dataset.  

\section{Interpretation}
\label{Sec:Interpretation}

\subsection{Millisecond Pulsars}
\label{MSPs}

Twenty-five MSPs have been phase-resolved in 47 Tuc \cite{freire}. The properties of these MSPs at $\gamma$-ray and radio wavelengths appear compatible with the MSPs in our local neighbourhood \cite{47tucfermiscience, freire} as is the range of X-ray luminosities \cite{chandra1,chandra2}. We therefore assume that the MSPs within 47 Tuc have similiar $\gamma$-ray properties to our local MSPs and thus model the MSPs in 47 Tuc with the same spectral shape as the \textit{Fermi}-LAT detected local MSPs. In particular, we use the spectral model derived by \citet{xingwang}, who stacked the \textsc{pass8} $0.1-300$ GeV spectra of 39 (out of the 40) MSPs reported in the second LAT pulsar catalog (2PC; \cite{2pc}). The best-fit spectral shape was a power-law with an exponential cut-off, a spectral index of $\Gamma=1.54^{+0.10}_{-0.11}$ and a cut-off energy of $E_c = 3.70^{+0.95}_{-0.70}$ GeV \cite{xingwang}. 



The normalisation of this spectral fit is related to the number of $\gamma$-ray bright MSPs in 47 Tuc. Since the angular resolution of \textit{Fermi}-LAT does not allow us to resolve individual MSPs within 47 Tuc, we initially consider the normalisation of the MSP spectral fit to be a free parameter. This conservative approach allows us to account for all MSPs within 47 Tuc, even those below the detection sensitivity threshold of the \textit{Fermi}-LAT. In parallel with this, we considered a second scenario where we determine the number of $\gamma$-ray bright MSPs by considering \textit{Chandra} X-ray observations, which have resolved 23 MSPs within 47 Tuc \cite{chandra1, chandra2, chandra3}. The $\gamma$-ray flux expected from these 23 X-ray bright MSPs was calculated using the X-ray-to-$\gamma$-ray MSP flux ratio of $\langle$log$(G_{100}/F_X)\rangle = 2.31$, derived in the 2PC \cite{2pc}. The combined flux from all 23 MSPs was then used to fix the normalisation of the MSP spectral fit.


\subsection{Dark Matter}

At present there is no evidence for DM halos in globular clusters \cite{HeggieHut1996,MashchenkoSills2005,Bradford2011,lane}. Globular clusters may have been originally embedded in dark halos, subsequently destroyed by stellar dynamical heating and tidal disruption by the host galaxy. However, the evidence for an IMBH within 47 Tuc \cite{imbh} leads us to rethink this common picture. In particular, depending on the formation of the central IMBH and the dynamical evolution of 47 Tuc, the cluster may have retained a sharply peaked DM distribution in its inner regions, which would enhance DM annihilation rates and related $\gamma$-ray fluxes. Therefore, we investigate whether the observed $\gamma$-ray emission from 47 Tuc can, in part, be attributed to a spiky DM distribution around the IMBH. To that end we search for components in the $\gamma$-ray spectrum of 47 Tuc that cannot be attributed to known astrophysical sources. This approach has already provided tantalizing hints of a DM population clustered around the supermassive black hole at the centre of the active galaxy Centaurus A \cite{brown}.

The adiabatic BH formation scenario would lead to an enhanced DM density, with a profile going as $r^{-\gamma_{\rm sp}}$ with $2.25 \leqslant\gamma_{\rm sp} \leqslant 2.5$, referred to as a DM spike \cite{gondolo}. However, adiabaticity is not guaranteed and several dynamical processes may have affected a spike, with an unclear outcome. For instance, dynamical relaxation with stars can lead to an equilibrium density profile $\propto r^{-3/2}$ \cite{Gnedin2004}. Due to the high density of stars in a globular cluster like 47 Tuc, this process is likely to have played a role. It is worth highlighting that the situation is different in a radio galaxy like Centaurus A \cite{brown} which is dynamically younger than a globular cluster.

As a result, to account for both the presence of an IMBH and dynamical effects, we assume the following DM profile:
\begin{align}
\label{DM_profile}
\rho(r) = 
\begin{cases}
0 & r < 2R_{\mathrm{S}} \\
\dfrac{\rho_{\rm sp}(r) \rho_{\rm sat}}{\rho_{\rm sp}(r) + \rho_{\rm sat}} & 2R_{\mathrm{S}} \leqslant r < R_{\rm sp} \\
\rho_0 \left( \dfrac{r}{R_{\rm sp}} \right)^{-5} & r \geqslant R_{\rm sp},
\end{cases}
\end{align}
where 
\begin{equation}
\rho_{\rm sp}(r) = \rho_0 \left( \dfrac{r}{R_{\rm sp}} \right)^{-3/2},
\end{equation}
and the effect of DM self-annihilation on the profile is accounted for via the saturation density
\begin{equation}
\rho_{\rm sat} = \dfrac{m_{\mathrm{DM}}}{\left\langle \sigma v \right\rangle t_{\mathrm{BH}}},
\end{equation}
with $m_{\mathrm{DM}}$ the mass of the DM candidate, $\left\langle \sigma v \right\rangle$ the velocity-averaged annihilation cross section, and $t_{\mathrm{BH}}$ the age of the central IMBH, which we take to be $\sim 11.75\, \rm Gyr$ \cite{imbh}. The radial extension of the spike $R_{\mathrm{sp}}$ should be of the order of the BH influence radius, $G M_{\mathrm{BH}}/\sigma_{*}^{2}$ \cite{peebles72}. The extended $M_{\mathrm{BH}}$-$\sigma_{*}$ relation for IMBHs \cite{tremaine2002} gives an estimated value of $\sigma_{*} \approx 10\ \rm km\ s^{-1}$ for the stellar velocity dispersion, and $R_{\mathrm{sp}} \approx 0.1\, \rm pc$ for M$_{\mathrm{BH}} = 2.3 \times 10^{3}\, $M$_{\odot}$. Outside of the spike, we assume the profile cuts off as $r^{-5}$ to keep a low DM content in the outer parts of the cluster. This cutoff is a priori ad hoc but could for instance originate from tidal stripping. The inner cutoff is related to capture of DM particles by the BH \cite{sadeghian2013}. The DM profile is normalized by requiring the mass inside the spike $M_{\mathrm{sp}}$ be of the order of the BH mass, which yields $\rho_{0} \approx (3-\gamma_{\mathrm{sp}}) M_{\mathrm{BH}}/(4 \pi R_{\mathrm{sp}}^{3})$. This gives a total DM mass in the cluster of about $4 \times 10^{3}\, $M$_{\odot}$, which is below 1\% of the total mass of 47 Tuc, and is therefore consistent with the low amount of DM favoured by the velocity dispersion profile of 47 Tuc \cite{lane}.

The resulting DM-induced $\gamma$-ray flux is given as usual by the volume integral of $\rho^{2}$:
\begin{equation}
\dfrac{\mathrm{d}n}{\mathrm{d}E_{\gamma}} = \dfrac{\left\langle \sigma v \right\rangle}{\eta m_{\rm DM}^{2} d^{2}} \dfrac{\mathrm{d}N_{\gamma}}{\mathrm{d} E_{\gamma}} \int_{0}^{R} \! \rho^{2}(r) r^2 \, \mathrm{d}r,
\end{equation}
where the distance to 47 Tuc is $d = 4\, \rm kpc$ \cite{McLaughlin2006,baumgardt2017} and the radial extension of the cluster is $R = 56\, \rm pc$ \cite{47tucdist}.\footnote{The precise value of the radial extension of the cluster is not relevant since we compute the DM-induced $\gamma$-ray flux for a sharply peaked profile.}

One may wonder whether the presence of a DM spike with a mass of the order of the IMBH mass would affect the kinematics of stars in the cluster. Detailed studies are required to quantify the effect of a spike on the kinematics of stars, but this is likely to require a much higher astrometric and spectroscopic precision. As of yet, the uncertainty on the mass of the IMBH from the study of Ref.~\cite{imbh} is quite large and leaves room for the DM component that we considered here.


To model the DM-induced contribution to the $\gamma$-ray spectrum of 47 Tuc, we considered prompt emission from DM annihilation into $b\bar{b}$ quarks, which gives a spectral shape that follows the trend of the data. The DM mass and annihilation cross section, $\langle\sigma v\rangle$, were treated as free parameters. 




\subsection{Results: combined model}

\begin{figure}[h!]
\includegraphics[width=0.9\columnwidth]{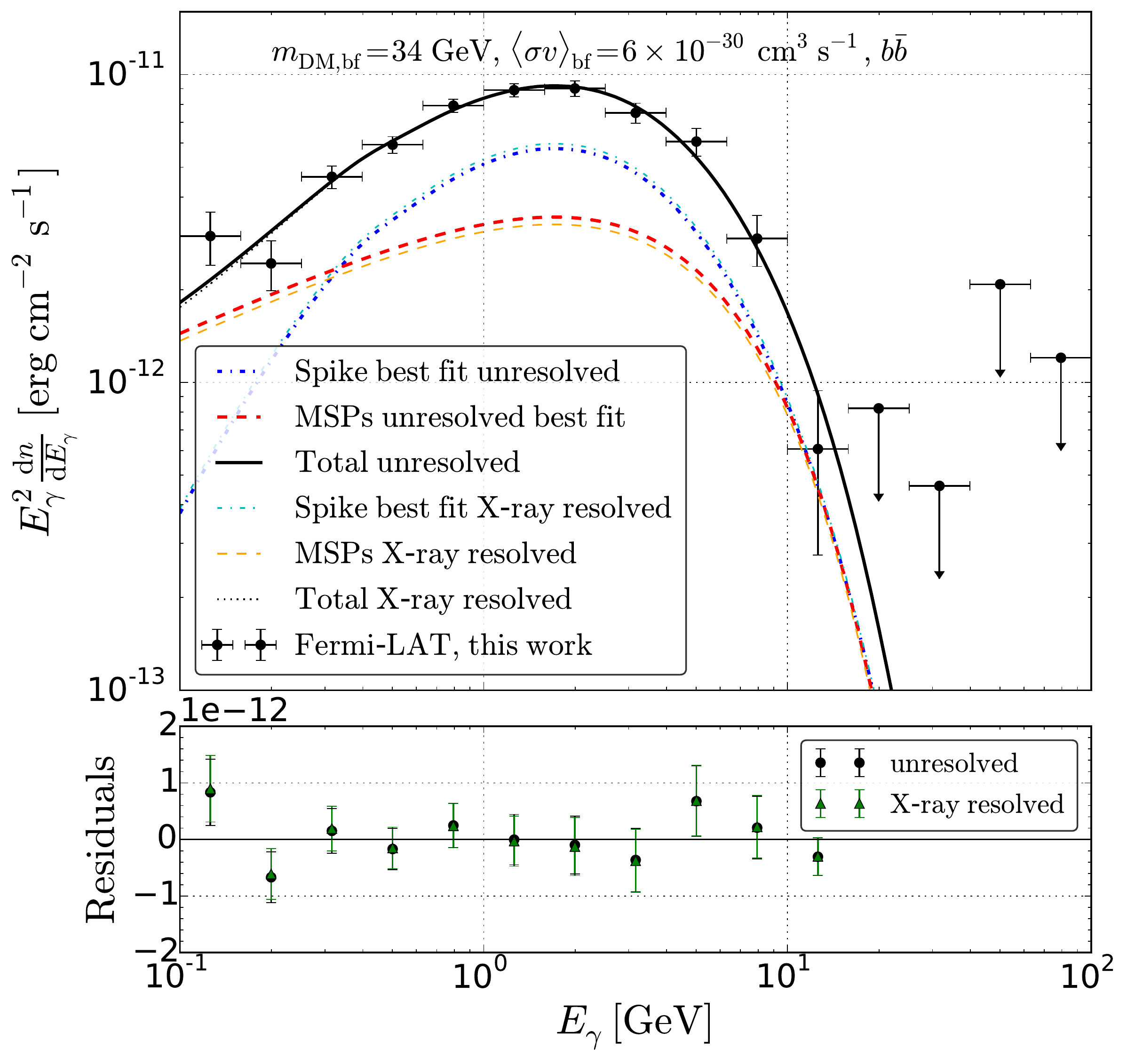}
\caption{Best `DM $+$ MSP' fit to 47 Tuc's $\gamma$-ray spectrum, for both methods for characterising the MSP contribution. DM annihilation via the $b\bar{b}$ channel, For both MSP scenarios considered, best-fit DM mass was found to be 34 GeV, with $\langle\sigma v\rangle \sim 6 \times 10^{-30}\, \rm cm^{3}\, s^{-1}$.}
\label{awesome2}
\end{figure}

We fit 47 Tuc's $\gamma$-ray spectrum with our `MSP$+$DM-spike' model by considering the two different scenarios for the MSP contribution described in Sec.~\ref{MSPs}. Assuming that the 23 known X-ray bright MSPs constitute all the MSPs present in 47 Tuc, using the average $\gamma$-ray$/$X-ray ratio to estimate the MSP $\gamma$-ray emission and leaving the DM mass and velocity averaged annihilation cross section free to vary, the best-fit solution has a DM mass of 34 GeV and a $\langle\sigma v\rangle$ of $6 \times 10^{-30}\, \rm cm^{3}\, s^{-1}$. The best-fit model and residuals can be seen in Fig.~\ref{awesome2}.

We note that the spike radius is degenerate with the annihilation cross-section. For instance, if we consider a more extended spike with $R_{\rm sp} = 1\, \rm pc$, the resulting best-fit cross section becomes $6 \times 10^{-27}\, \rm cm^{3}\, s^{-1}$. Moreover, without a spike, the annihilation cross-section required to produce a gamma-ray flux of the order of that observed is very large and already excluded by indirect searches. Finally, we have $\mathrm{d}n/\mathrm{d}E_{\gamma} \propto M_{\rm sp}^{2}$, where we took $M_{\rm sp} = M_{\rm BH}$, so the impact of changing the mass enclosed in the spike can be readily translated into a change in the best-fit cross section.


The second approach is a model-independent one and simply assumes that the MSP contribution is due to an ensemble of unresolved MSPs. As such, we leave the normalisation of the MSP spectral model free to vary. As before, we simultaneously fit the MSP and DM contributions, leaving the DM mass and velocity averaged annihilation cross section free to vary, assuming the enhanced density profile around the IMBH. The best-fit solution, assuming an unresolved population of MSPs, also has a DM mass of 34 GeV and a resultant $\langle\sigma v\rangle$ of $6 \times 10^{-30}\, \rm cm^{3}\, s^{-1}$. This fit, along with its residuals, can be seen in Fig.~\ref{awesome2}. It is interesting to note that when allowing for unresolved MSPs, the best-fit unresolved MSP flux is consistent with the flux expected from the 23 X-ray resolved MSPs.

\begin{figure}[h!]
\includegraphics[width=\columnwidth]{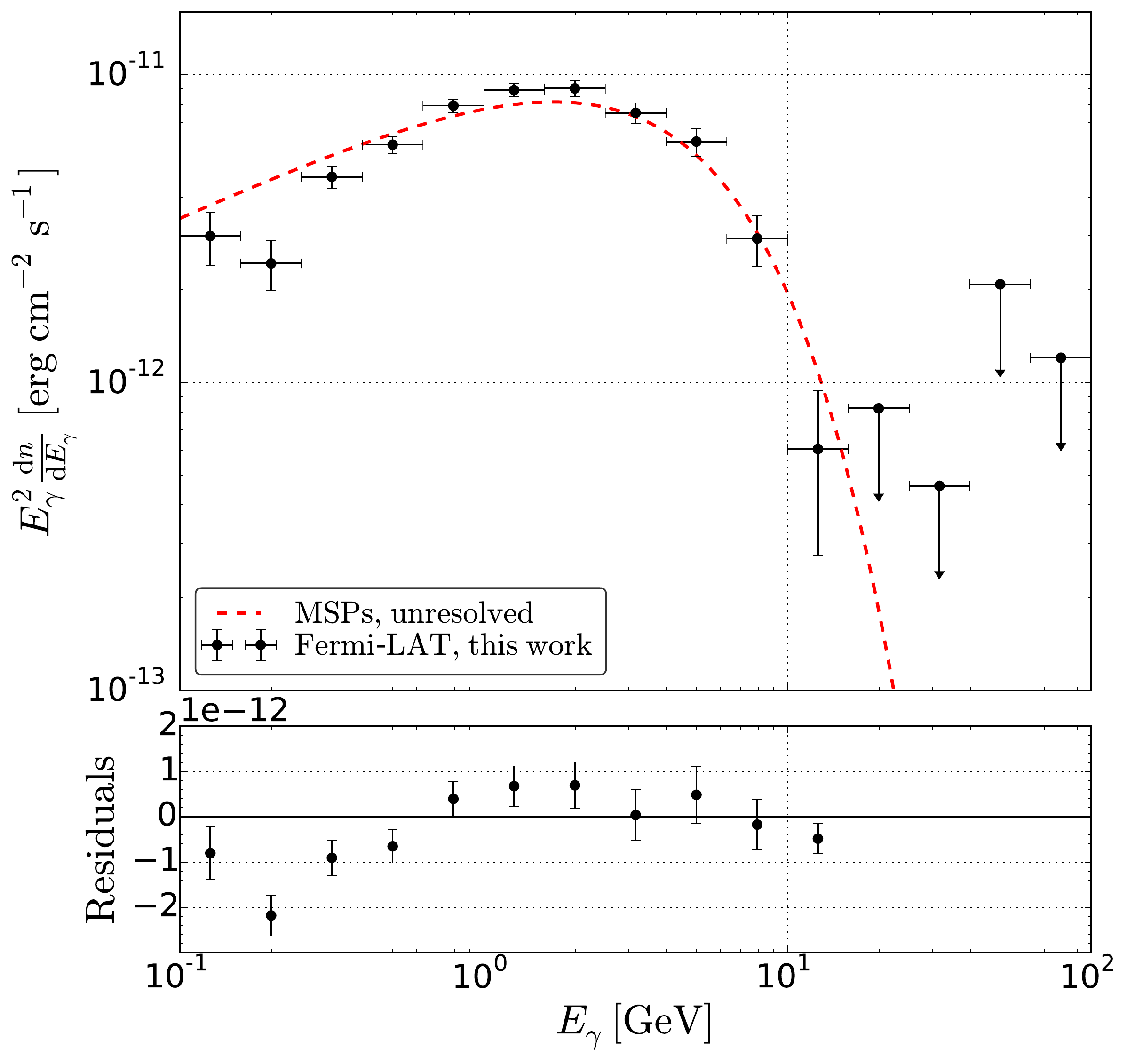}
\caption{Best `MSP-only' fit to 47 Tuc's spectrum, assuming that the $\gamma$-ray flux is solely due to a population of MSP. We describe the MSP population with the Xing \& Wang spectral shape derived by averaging over all MSPs in the 2PC. The residuals show systematic discrepancies between the model and the observed spectrum.}
\label{awesome1}
\end{figure}

To determine the significance of the DM contribution, we take the conservative approach and consider an `extreme' MSP situation whereby we attribute all of the observed $\gamma$-ray flux to a population of unresolved MSPs, with no DM contribution considered at all. For this, we used a maximum likelihood approach to fit 47 Tuc's spectrum with an `MSP-only' model, assuming the same spectral model as that of our local $\gamma$-ray bright MSPs, as derived by Xing \& Wang. The normalisation during the likelihood fit was left free to vary. The best-fit solution can be seen in Fig.~\ref{awesome1}, which shows systematic discrepancies between the model and the observed spectrum, especially below $E_{\gamma}\approx 0.8$~GeV. The log-likelihood value of this MSP-only fit was then compared to those of the `unresolved MSP $+$ DM' and `resolved MSP $+$ DM' fits, using the standard definition of TS as discussed in Sec.~\ref{Sec:Analysis}. The comparison finds both DM models are significantly preferred over the MSP-only model with a TS $=40$ for each, which equates to $>5\sigma$. We stress, however, that this significant preference does not rule out the MSP-only model, if $\gamma$-ray spectra of our local MSP population are not representative of those in 47 Tuc.


\section{Discussion}
\label{Sec:Discussion}

With the significance of our DM$+$MSP model confirmed, we turn our attention to exploring possible $\gamma$-ray sources other than DM annihilation. One such is magnetospheric emission from matter around 47 Tuc's IMBH \cite{magneto}; however, given that this emission is expected to be proportional to the black hole mass \cite{brownngc1275}, such emission would be below the sensitivity of \textit{Fermi}-LAT. Furthermore, there is no radio or X-ray evidence of the IMBH accreting, suggesting that there are insufficient particles for magnetospheric emission \cite{derijcke,grindlay}.

Other possible sources of $\gamma$-rays in 47 Tuc are LMXBs and Cataclysmic Variables (CVs). While CVs were discovered to be $\gamma$-ray bright by the \textit{Fermi}-LAT \cite{cv1,cv2,cv3}, the $\gamma$-ray emission is transient, on the timescale of days, and only CVs in our local Galactic neighbourhood have been observed to be $\gamma$-ray bright, suggesting that their $\gamma$-ray flux is observed simply because of their proximity to us \cite{ourCV}. LMXBs are not known to be $\gamma$-ray emitters; all confirmed $\gamma$-ray emitting binary systems are wind-driven systems and are classed as high-mass X-ray binaries. 

Finally, we turn our attention to the assumption that MSPs within 47 Tuc have the same $\gamma$-ray spectral shape as those of the $\gamma$-ray bright MSPs in our local Galactic neighbourhood. Throughout our studies, we have modeled the 47 Tuc's MSPs with the spectral shape derived by Xing \& Wang \cite{xingwang}. Xing \& Wang determined their spectral shape by individually analysing 7.5 years of observations for all MSPs in the 2PC, stacking the resultant spectra and deriving an average spectrum. All MSPs in the 2PC reside in our local Galactic neighbourhood and one may question whether these MSPs are representative of those found in globular clusters. Previous radio, X-ray and $\gamma$-ray studies have found no difference in observational properties such as luminosity, spectral shape and cut-off energy, nor in intrinsic MSP properties such as `spin-down to $\gamma$-ray' luminosity conversion and pulsar timing, between local MSPs which make up the 2PC, and those in globular clusters \cite{derijcke,grindlay,47tucfermiscience,fermiGCs}. Observational evidence therefore suggests that the $\gamma$-ray spectra of MSPs in 47 Tuc are not significantly different to those of the local MSPs. 

Nonetheless, relaxing our assumption that the average spectral shape of the MSPs within 47 Tuc is the same as our local $\gamma$-ray bright MSPs, we use a maximum likelihood approach to fit 47 Tuc with a general exponential cut-off power-law model, leaving the spectral index, cut-off energy and normalisation free. The resultant best-fit has a spectral index $\Gamma= 1.21 \pm 0.06$ and a cut-off energy of $E_c = 2.4 \pm 0.2$ GeV. While there is no significant difference between this fit when compared to the `MSP$+$DM' fit, the spectral index is harder, and the power cuts-off at a lower energy when compared to our local MSPs. Importantly, there is no evidence that 47 Tuc's $\gamma$-ray emission is dominated by a few bright MSPs \cite{47tucfermiscience}, meaning that the $\gamma$-ray spectrum cannot be explained by a few MSPs with vastly different $\gamma$-ray properties to our local MSPs. As such, this suggests that if 47 Tuc's $\gamma$-ray emission is solely due to MSPs, the $\gamma$-ray properties of these MSPs appear to be markedly different to those of our local Galactic neighbourhood, which is not supported by observational results at X-ray and radio wavelengths. 

In both MSP scenarios we have considered, the addition of a DM component, with an enhanced density around 47 Tuc's IMBH, results in a significant improvement to the spectral fit of 47 Tuc at the level of TS $=40$ ($>5\sigma$). Interestingly, for both MSP scenarios considered, the DM mass is found to be 34 GeV, which is essentially the same as the best-fit DM explanation for the Galactic centre ``excess" when assuming DM annihilation into $b$ quarks \cite{gordon,hooper}. However, the value of our best-fit annihilation cross section is too small to account for the observed cosmological DM abundance, but this might a hint for a rich dark sector with several (nonthermal) DM candidates \cite{boehm2004,zurek2009}, or a combination of velocity-dependent and independent contributions to the annihilation cross section.

As mentioned in Sec.~\ref{Sec:Interpretation}, we have assumed that a DM population is in addition to the $\sim2300$ M$_{\astrosun}$ IMBH within 47 Tuc, and have not considered the possibility that the inferred IMBH itself is DM, although this is very unlikely since it would require extreme clustering of DM in the central regions, with a DM mass of the order of the IMBH. This would not be expected without a black hole already present at the centre.

\section{Conclusions}
\label{Sec:Conclusions}

In this paper we report our study of 47 Tuc using 9 years of \textit{Fermi}-LAT \textsc{pass8} observations. We find that 47 Tuc's observed $\gamma$-ray flux cannot solely be explained by a population of MSPs due to systematic discrepancies between the predicted and observed flux. Motivated by the recent evidence of an IMBH within 47 Tuc, we model 47 Tuc's 9-year spectrum with two sources of $\gamma$-rays: MSPs and DM with an enhanced density around the IMBH. For the MSP description, we consider both the resolved and unresolved MSP population within 47 Tuc, with remarkable agreement between the two suggesting that all MSPs within 47 Tuc are resolved. For either MSP description, a maximum likelihood analysis reveals that a DM component is significantly preferred, $>5\sigma$, when compared to a MSP only explanation.  As such, our work constitutes the first possible evidence of DM within a globular cluster. This could have important consequences when trying to understand how globular clusters formed, or signifies that 47 Tuc is a unique globular cluster.

\section{Acknowledgements} The authors would like to thank the referees for their comments which have undoubtedly improved the quality and clarify of the paper. We thank Joe Silk, Cameron Rulten and Jamie Graham for helpful discussions. We acknowledge the excellent data and analysis tools provided by the \textit{Fermi}-LAT collaboration. AMB and PMC acknowledge the financial support of the UK Science and Technology Facilities Council consolidated grant ST/P000541/1. TL receives financial support from CNRS-IN2P3 and acknowledges support from the European Unions Horizon 2020 research and innovation program under the Marie Sk\l{}odowska-Curie grant agreement Nos. 690575 and 674896; beside recurrent institutional funding by CNRS-IN2P3 and the University of Montpellier. This research was supported in part by Perimeter Institute for Theoretical Physics. Research at Perimeter Institute is supported by the Government of Canada through Industry Canada and by the Province of Ontario through the Ministry of Economic Development and Innovation. 

\end{document}